\documentclass[12pt]{article}
\usepackage{latexsym,epsfig,amssymb, amsmath, cite, mcite}
\usepackage[utf8]{inputenc}

\begin{document}

\begin{titlepage}

\begin{center}
\footnotesize{\it{Barcelona Postgrad Encounters on Fundamental Physics}}
\end{center}

\bigskip

\begin{center}

\vskip 2cm

{\LARGE \bf The many surprises of \\[3mm]maximal supergravity} \\

\vskip 1.0cm

{\bf Andrea Borghese$^1$, Adolfo Guarino$^2$, Diederik Roest$^1$}\\

\vskip 0.5cm

{\em $^1$ Centre for Theoretical Physics,\\
University of Groningen, \\
Nijenborgh 4, 9747 AG Groningen, The Netherlands\\
{\small {\tt \{ a.borghese, d.roest \} @rug.nl}}} \\

\vskip 0.3cm

{\em $^2$ Albert Einstein Center for Fundamental Physics,\\
Institute for Theoretical Physics, Bern University,\\
Sidlerstrasse 5, CH3012 Bern, Switzerland\\
{\small {\tt guarino@itp.unibe.ch}}}

\end{center}

\vskip 2cm

\begin{center} {\bf ABSTRACT}\\[3ex]

\begin{minipage}{13cm}
\small

We describe recent developments regarding gauged $\mathcal{N}=8$ supergravity in $D=4$. Using the embedding tensor formulation we show how to classify all the extrema of this theory with a ${\rm{G}}_2$ residual gauge symmetry. Our classification contains all the vacua of the recently discovered \cite{Dall'Agata:2012bb} family of ${\rm{SO}}(8)$ gauged maximal supergravities.

\end{minipage}

\end{center}

\vspace{2cm}

\flushleft{\footnotesize{\it{Prepared for Barcelona Postgrad Encounters on Fundamental Physics,\\ 17-19 October 2012, Barcelona, Spain.}}}

\vfill

\end{titlepage}


\tableofcontents

\section{Introduction}
Maximal supergravities in four dimensions have been discovered at the end of the se\-venties and, after more than thirty years, still have surprises in store. \\
The ungauged version \cite{Cremmer:1979up} can be obtained by torus reduction of eleven dimensional supergravity. Many distinctive features already show up at this level. In particular the field content is completely fixed and consists of the $\mathcal{N} = 8$ supergravity multiplet alone without any possibility of coupling matter multiplets. The theory contains 70 real scalar fields which can be viewed as coordinates of an ${\rm{E}}_{7}/{\rm{SU}}(8)$ coset manifold. During the last decade ungauged maximal supergravity has received some attentions because of its special ultraviolet behaviour. Four-graviton scattering amplitudes are finite up to four loops \cite{Bern:2011qn}. This could be a hint of the fact that the theory is perturbatively finite providing the first and only instance of an ultraviolet finite, point-like theory of quantum gravity. \\
In a gauged supergravity some vectors in the spectrum are used to gauge a subgroup of the duality group ${\rm{E}}_{7}$. The gauging procedure is the only known consistent deformation of maximal supergravity. The main advantage of carrying out this procedure is the introduction of a potential which enriches the dynamics in the scalar sector. \\
The first example of this class of maximal supergravities is the ${\rm{SO}}(8)$ gauged theory \cite{deWit:1982ig} in which all the 28 physical vectors in the spectrum are used to gauge an ${\rm{SO}}(8)$ subgroup of ${\rm{E}}_{7}$. This gauged theory can be obtained by sphere reduction of eleven dimensional supergravity and has been intensively studied from many perspectives. Its vacuum structure has been analysed restricting to truncated sectors \cite{Warner:1983vz} and, in recent years, a scan of all vacua having cosmological constant within a certain range of values has been performed using numerical techniques \cite{Fischbacher:2011jx}. With the advent of the gauge/gravity duality the ${\rm{SO}}(8)$ gauged maximal supergravity has been used to construct models of holographic superconductivity \cite{Donos:2011ut, Bobev:2010ib, Fischbacher:2010ec, Bobev:2011rv}. Furthermore the theory should be dual to the ABJM three dimensional CFT \cite{Aharony:2008ug}. \\
Some other gauged maximal supergravities have been constructed with a variety of gauge groups \cite{Hull:1984ea, Hull:2002cv}. Theories with a compact gauge group usually displays AdS vacua while, whenever a non-compact group is gauged, it is possible to find dS critical points though perturbatively unstable\footnote{dS/Minkowski critical points are perturbatively unstable whenever, expanding around those points, some scalar excitation has a negative squared mass. For AdS critical point, a scalar field is unstable if its mass squared is below the Breitenlohner-Freedman (BF) bound \cite{Breitenlohner:1982bm}. In our conventions, instability means $m^2 L^2 = 3 m^2 / |V| \leq - 9/4$.}. The quest for a stable dS vacuum in maximal supergravity has been, so far, inconclusive\footnote{However see \cite{Dall'Agata:2012sx} for the last interesting development in this direction.}. Either this difficulty could be overcome by spotting the right gauging or it could be a hint of a fundamental obstruction against the realisation of dS vacua in maximal supergravity. \\
Gauged supergravity plays also a role in the field of flux compactifications. As described above, some gaugings could be obtained by compactifications of higher dimensional theories. Other gaugings do not have a higher dimensional origin but show interesiting features. it is worth understanding which higher dimensional ingredients (compactification manifold, fluxes or extended sources such as branes or orientifold planes) need to be used to obtain a certain gauged supergravity. \\
Due to the web of open questions and interesting applications it is worthwhile investigating the vacua of gauged maximal supergravity. This could give the possibility of answering some of those questions or provide models for holography. Our contribution takes a step in this direction. \\
Despite the fact that only a handful of gaugings have been worked out explicitly, in the last decade a gauging independent formulation of the theory \cite{deWit:2005ub, deWit:2007mt} has been constructed. This formulation makes use of an object called embedding tensor which allows to keep implicit the gauge group and still write down a Lagrangian as we will explain in section 2. \\
The embedding tensor formulation also allows to use group theory as a powerful tool to understand the structure of gauged supergravity. An example of this is the new family of ${\rm{SO}}(8)$ gauged supergravity discovered in \cite{Dall'Agata:2012bb}. Up to 2012, the ${\rm{SO}}(8)$ gauged theory was believed to be unique. Using group theory it is instead easy to show that there is a one parameter family of such theories with a different vacuum structure. This finding will be described in section 3. \\
Another interesting development which gives the possibility of understanding more clearly the vacuum structure of gauged supergravity is what we will call the ``going to the origin'' (GTTO) approach \cite{Dibitetto:2011gm, DallAgata:2011aa}. We will describe this approach in section 4 and show how it allows to classify exhaustively vacua preserving a given amount of symmetry. \\
In section 5 we will go through the example of critical points with a residual ${\rm{G}}_{2}$ symmetry \cite{Borghese:2012qm} and explain the relation between our findings and the new family of ${\rm{SO}}(8)$ maximal supergravity found in \cite{Dall'Agata:2012bb}.

\section{Embedding tensor formulation}

The field content of $\mathcal{N}=8$ supergravity in $D=4$ consists of a graviton, 8 gra\-vitini, 28 vectors, 56 spin $1/2$ fields and 70 scalar fields. In this section we briefly review the most recent construction of the gauged theory with the aim of underlining some fundamental concepts. We refer to \cite{deWit:2005ub, deWit:2007mt} for a more detailed and explicit treatment. \\
The 70 scalar fields are associated to isometries of the scalar manifold ${\rm{E}}_{7}/{\rm{SU}}(8)$. This means they are in a one to one correspondence with the 70 non compact generators of ${\rm{E}}_{7}$. We denote the adjoint of ${\rm{E}}_{7}$ with indices $\alpha , \beta \in \bf{133}$. The generators ${\rm{t}}_{\alpha}$ split in 63 compact and 70 non compact ones. \\
The theory construction relies on the ${\rm{E}}_{7}$ duality group. The 28 physical vectors, which we call electric, are supplemented with their magnetic counterpart and together they sit in the fundamental representation of the duality group. We write them as $A_{\mu}{}^{\mathcal{M}}$ with $\mathcal{M} \in \bf{56}$ of ${\rm{E}}_{7}$. Some of these vectors are used to gauge a subgroup ${\rm{G}}_{g} \subset {\rm{E}}_{7}$. The structure of derivatives on a generic field is hence modified according to
\begin{equation} \label{Derivatives}
\mathcal{D}_{\mu} = \partial_{\mu} - A_{\mu}{}^{\mathcal{M}} \, {\rm{\Theta}}_{\mathcal{M}}{}^{\alpha} \, {\rm{t}}_{\alpha} \, .
\end{equation}
The object denoted with $\rm{\Theta}$ is called embedding tensor. It selects the linear combination of vectors which is used to gauge a particular isometry of the scalar manifold. We can look at it as a set of charges which transform non trivially under the action of the duality group. In particular $\rm{\Theta} \in \bf{56} \times \bf{133}$ of ${\rm{E}}_{7}$. \\
In order for the gauging procedure to be consistent we need to impose some constraints on the embedding tensor: a set of linear constraints (LC) and a set of quadratic ones (QC). The LC are necessary to preserve invariance of the action under supersymmetry. They amount to ask the embedding tensor to belong to a single representation of ${\rm{E}}_{7}$
\begin{equation} \label{LC}
{\rm{\Theta}} \in \bf{912} \subset \bf{56} \times \bf{133} \, .
\end{equation}
The QC are instead necessary for the consistency of the gauging. In fact it is of fundamental importance that the charges of a gauge theory be invariant under the action of the gauge group. In our case the gauge group is a subgroup of the duality group. Thus we have to impose explicitly that, while transforming under ${\rm{E}}_{7}$, the embedding tensor be invariant under the action of ${\rm{G}}_{g} \subset {\rm{E}}_{7}$. This could be written in the following form
\begin{equation} \label{QC}
{\rm{\Theta}}_{\mathcal{M}}{}^{\alpha} \, {\rm{\Theta}}_{\mathcal{N}}{}^{\beta} \, {\rm{\Omega}}^{\mathcal{M} \mathcal{N}} = 0 \, ,
\end{equation}
where ${\rm{\Omega}}$ is antisymmetric and invariant under the action of ${\rm{E}}_{7}$. If we define an element of the gauge algebra with ${\rm{X}}_{\mathcal{M}} = {\rm{\Theta}}_{\mathcal{M}}{}^{\alpha} \, {\rm{t}}_{\alpha}$ and the generalised structure constants with ${\rm{X}}_{\mathcal{M} \mathcal{N}}{}^{\mathcal{P}} = {\rm{\Theta}}_{\mathcal{M}}{}^{\alpha} \, [ {\rm{t}}_{\alpha} ]_{\mathcal{N}}{}^{\mathcal{P}}$, equation (\ref{QC}) implies
\begin{equation} \label{Gauge algebra}
\big[ {\rm{X}}_{\mathcal{M}}, {\rm{X}}_{\mathcal{N}} \big] = - {\rm{X}}_{\mathcal{M} \mathcal{N}}{}^{\mathcal{P}} \, {\rm{X}}_{\mathcal{P}} \, ,
\end{equation}
where with $[ {\rm{t}}_{\alpha} ]_{\mathcal{N}}{}^{\mathcal{P}}$ we mean that the generator is in the ${\rm{E}}_{7}$ fundamental representation. Equation (\ref{Gauge algebra}) implies the closure of the gauge algebra. \\
Once embedding tensor components are chosen satisfying the LC and the QC, the gauge group is fixed and a consistent ${\rm{G}}_{g}$ gauged supergravity is obtained. Nevertheless it is not compulsory to specify the embedding tensor components. The gauged Lagrangian in \cite{deWit:2005ub, deWit:2007mt} is perfectly consistent and invariant under local supersymmetry transformations even if the embedding tensor is left unspecified but satisfies the constraints (\ref{LC}, \ref{QC}). In this sense the construction is gauging independent and gives the universal Lagrangian of four dimensional maximal supergravity. \\
We will give now some more detail regarding the scalar sector of the theory. Scalar fields will be denoted with $\vec{\phi}$. A central role in the theory construction is played by the scalar manifold coset representative. We will denote it with $\mathcal{V}(\vec{\phi})_{\mathcal{M}}{}^{\underline{\mathcal{M}}}$. This object is an element of the ${\rm{E}}_{7}$ group. We can act on it with a global ${\rm{E}}_{7}$ transformation from the left (on the index $\mathcal{M}$) and with local ${\rm{SU}}(8) \subset {\rm{E}}_{7}$ transformations from the right (on the index $\underline{\mathcal{M}}$). it is used to couple gauge field strength (which have global indices) to fermionic degrees of freedom (which have ${\rm{SU}}(8)$ local indices being ${\rm{SU}}(8)$ the R-symmetry group). From the coset representative $\mathcal{V}$ it is possible to define a metric on the scalar manifold $\mathcal{M} = \mathcal{V} \, \mathcal{V}^{T}$ and hence the scalar potential in its gauging independent form
\begin{equation} \label{Scalar potential E7}
V(\vec{\phi}, {\rm{\Theta}}) = \frac{1}{672} \big( {\rm{X}}_{\mathcal{M} \mathcal{N}}{}^{\mathcal{R}} \, {\rm{X}}_{\mathcal{P} \mathcal{Q}}{}^{\mathcal{S}} \, \mathcal{M}^{\mathcal{M} \mathcal{P}} \, \mathcal{M}^{\mathcal{N} \mathcal{Q}} \, \mathcal{M}_{\mathcal{R} \mathcal{S}} + 7 \, {\rm{X}}_{\mathcal{M} \mathcal{N}}{}^{\mathcal{Q}} \, {\rm{X}}_{\mathcal{P} \mathcal{Q}}{}^{\mathcal{N}} \, \mathcal{M}^{\mathcal{M} \mathcal{P}} \big) \, .
\end{equation}
it is worth pointing out some features of equation (\ref{Scalar potential E7}). First of all it is invariant under duality transformations in the following sense: for every transformation ${\rm{U}}$ acting on the scalars through the coset representative ${\rm{U}}_{\mathcal{M}}{}^{\mathcal{N}} \, \mathcal{V}(\vec{\phi})_{\mathcal{N}}{}^{\underline{\mathcal{N}}} = \mathcal{V}(\vec{\phi}')_{\mathcal{M}}{}^{\underline{\mathcal{N}}}$, there is a compensating transformation on the embedding tensor $\rm{U} \, \Theta = \Theta'$ such that $V(\vec{\phi}', {\rm{\Theta}}) = V(\vec{\phi}, {\rm{\Theta}}')$. Furthermore the scalar potential has in general a complicated non linear dependence on the scalar fields while it is homogeneous and quadratic in the embedding tensor components through the generalised structure constants. These considerations will prove useful in the next sections.

\section{The new family of ${\rm{SO}}(8)$ gaugings}
We will now specify the gauge group to be ${\rm{SO}}(8)$ and show how the family of ${\rm{SO}}(8)$ gauged maximal supergravities comes out from group theoretical considerations. We follow the argument of \cite{Dall'Agata:2012bb} in the derivation. \\
As explained in the previous section, the embedding tensor components should be singlets under the gauge group. Thus we need to consider the branching of the $\bf{912}$ representation of ${\rm{E}}_{7}$ duality group under ${\rm{SO}}(8)$
\begin{equation} \label{912 branching under SO8}
{\bf{912}} = 2 \times \big( {\bf{1}} + {\bf{35}}_{v} + {\bf{35}}_{s} + {\bf{35}}_{c} + {\bf{350}} \big) \, ,
\end{equation}
where the $\bf{35}$ representations are related to the vector, spinor and conjugate spinor representation of ${\rm{SO}}(8)$. In there we find two singlets and the parameter interpolating between these singlets span the family of ${\rm{SO}}(8)$ gauged theories. Despite the simplicity of this argument it took thirty years to acknowledge that in fact there is not just a single theory but rather a one parameter family of different theories. \\
At this point we introduce a bit of notation. We consider the ${\rm{SL}}(8)$ symplectic frame \cite{deWit:2005ub} in which the 56 vectors can be split in 28 electric ones transforming in the $\bf{28}$ of ${\rm{SL}}(8)$ and 28 magnetic transforms in the contravariant representation, the $\bf{28}'$. We denote vectors as $A_{\mu}{}^{\mathcal{M}} = \{ A_{\mu}{}^{[AB]}, A_{\mu \, [AB]} \}$ with $A,B \in {\bf{8}}$ of ${\rm{SL}}(8)$. The ${\rm{SO}}(8)$ group is trivially embedded in ${\rm{SL}}(8)$ and the two singlets in (\ref{912 branching under SO8}), coupled to electric and magnetic vectors respectively, can be written as
\begin{equation}
{\rm{\Theta}}_{\mathcal{M}}{}^{\alpha} = \big\{ {\rm{\Theta}}_{AB}{}^{C}{}_{D}, \, {\rm{\Theta}}^{ABC}{}_{D} \big\} = \big\{ \cos \omega \, \delta^{C}_{[A} \, \delta_{B]D} , \, \sin \omega \, \delta_{D}^{[A} \, \delta^{B]C} \big\} \, .
\end{equation}
As already explained, the $\omega$ parameter interpolates between electric and magnetic components of the embedding tensor. \\
For different values of $\omega$ we are in front of different maximal supergravity theories with ${\rm{SO}}(8)$ gauge group. In order to show this and make contact with the next sections we consider a truncation of these theories: the restriction to the ${\rm{G}}_2$-invariant sector \cite{Borghese:2012zs}. As the name suggests, the truncation contains all fields in the spectrum which are invariant under a subgroup of the duality group, namely the ${\rm{G}}_2$ term in 
\begin{equation} \label{from E7 to G2}
{\rm{E}}_{7} \supset {\rm{SL}}(2) \times {\rm{G}}_{2} \, ,
\end{equation}
while other fields are set to zero. The truncation is $\mathcal{N} = 1$ supersymmetric, in the bosonic sector there are no vectors and only one complex field $z$ is retained. The scalar potential (\ref{Scalar potential E7}) reduces to
\begin{equation} \label{potential G2}
V = e^{\mathcal{K}} \big[ \mathcal{K}^{z \bar{z}} \, \big( \mathcal{D}_{z} \mathcal{W} \big) \, \big( \mathcal{D}_{\bar{z}} \overline{\mathcal{W}} \big) - 3 \, \mathcal{W} \, \overline{\mathcal{W}} \big] \, ,
\end{equation}
with
\begin{equation} \label{super and Kahler potentials}
\mathcal{K} = -7 \, \ln \big[ -1 + \dfrac{1}{1+z} + \dfrac{1}{1+\bar{z}} \big] \, , \quad \mathcal{W} = \dfrac{\sqrt{2}}{(1+z)^{7}} \big[ \big( 1 + 7 \, z^{4} \big) \, e^{i \, \omega} + \big( 7 \, z^{3} + z^{7} \big) \, e^{- i \, \omega} \big] \, .
\end{equation}
The plots in Figure 1 show the form of the scalar potential in terms of two real fields $\{\phi_{1} , \phi_{2}\}$ with $z = -(\phi_{1} + i \, \phi_{2})$ for two distinct values $\omega = 0, \pi/8$.
\begin{figure}[h!]
\begin{center}
\includegraphics[scale=.7]{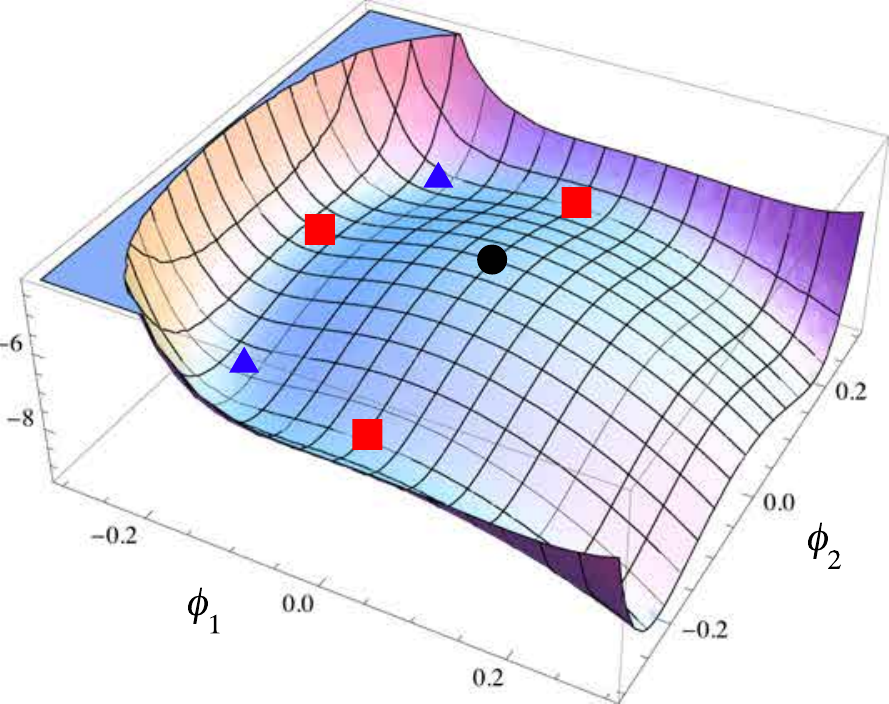} \hspace{0.5cm}
\includegraphics[scale=.7]{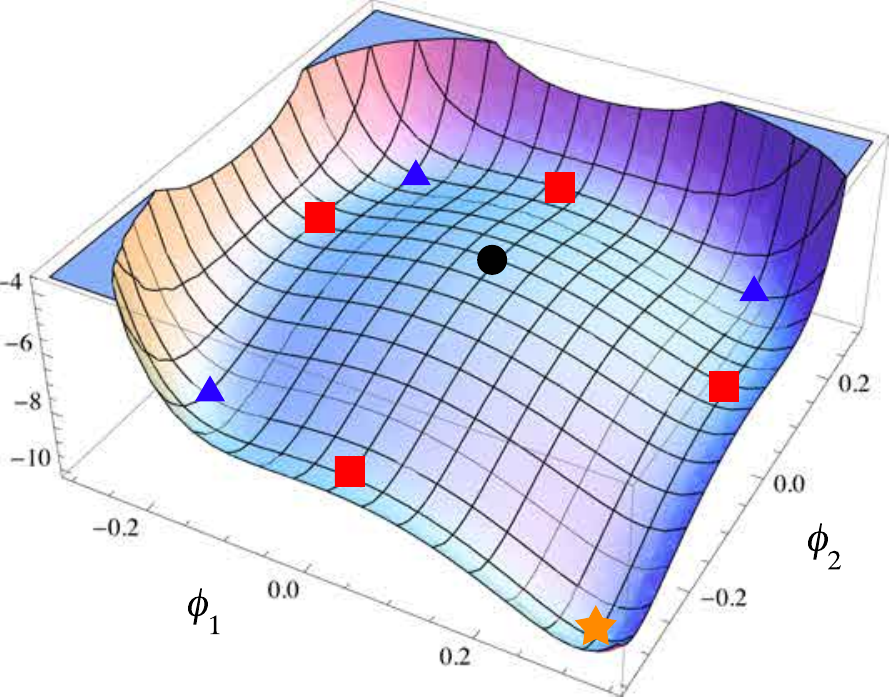}
\caption{Scalar potential of the new ${\rm{SO}}(8)$ gauged maximal supergravities in the ${\rm{G}}_{2}$ invariant sector. On the left $\omega=0$ while on the right $\omega=\pi/8$.}
\label{NewSO8SUGRAomega0}
\vspace{-0.8cm}
\end{center}
\end{figure} \\
The left plot contains six critical points divided in three different groups. At the origin there is an $\mathcal{N}=8$ supersymmetric critical point preserving the full gauge symmetry group ${\rm{SO}}(8)$ marked with a black circle. The three points marked with red squares are non supersymmetric critical points in which the gauge group is broken from ${\rm{SO}}(8)$ to ${\rm{SO}}(7)$. Finally the blue triangles are $\mathcal{N}=1$ supersymmetric critical points preserving a residual ${\rm{G}}_2$ symmetry. \\
Three new critical points appear in the right plot of Figure 1. Two of them fall in the same groups which were present in the $\omega = 0$ case. The last one, marked with an orange star, preserves again a ${\rm{G}}_2$ symmetry but is non supersymmetric. \\
These plots show that for different values of $\omega$ the potential (\ref{potential G2}) displays a different vacuum structure. In particular, the appearence of a new type of critical point (the non supersymmetric ${\rm{G}}_2$) demonstrates that the two theories at $\omega = 0$ and $ \pi/8$ are fundamentally different. \\
Even though the vacuum positions and the cosmological constant values change with $\omega$, the normalised squared masses $m^{2} L^{2}$ of the two scalar excitations remain the same for critical points preserving the same amount of supersymmetry and residual gauge symmetry. In particular, for the maximally supersymmetric critical point we have $m_{1,2}^{2} L^{2} = -2$. For the $\mathcal{N}=1$, ${\rm{G}}_2$ symmetric ones we have $m_{1,2}^{2} L^{2} = 4 \pm \sqrt{6}$. For the non supersymmetric ${\rm{SO}}(7)$ invariant critical points we have $m_{1}^{2} L^{2} = 6$ and $m_{2}^{2} L^{2}= -6/5$. Finally for the $\mathcal{N}=1$, ${\rm{G}}_2$ symmetric one we have $m_{1,2}^{2} L^{2}= 6$. \\
There is a motivation for the fact that moving in the $\omega$ space we find different theories. To explain that we need to go back to the definition of duality group. Such a group is defined as the invariance group of the set of equations of motions (EOM) plus Bianchi identities (BI) of a supergravity theory. Acting with the duality group we do change the Lagrangian of the theory but we do not change the set of EOM plus BI thus leaving unchanged physical observables (such that critical points). Hence it is clear that an $\omega$ transformation must sit outside ${\rm{E}}_{7}$. Indeed they belong to the complement of ${\rm{E}}_{7}$ inside ${\rm{Sp}}(56) \supset {\rm{E}}_{7}$. The symplectic group ${\rm{Sp}}(56)$ is the electro-magnetic (E-M) duality group rotating among each others electric and magnetic vectors \cite{Dall'Agata:2012bb}. 

\section{Going to the origin}
In this section we will describe an alternative approach for the study of vacuum structure in supergravity \cite{Dibitetto:2011gm}. 
The GTTO approach relies on the special form of extended supergravity scalar manifolds. In particular for (half)-maximal supergravities these are always homogeneous coset manifold ${\rm{G/H}}$. For such spaces there is no preferred point due to the transitive action of ${\rm{G}}$ on the elements of the manifold. A way of rephrasing this sentence is stating that every point of the coset can be transformed to the origin\footnote{By definition, the origin of the coset space in a given parametrisation is identified with the point $\vec{\phi} = 0$.} of the coset. \\
In the origin of moduli space we loose covariance w.r.t. the full ${\rm{E}}_{7}$ group and we are left with ${\rm{SU}}(8)$ covariant objects. Every tensor in a given ${\rm{E}}_{7}$ representation branches in different ${\rm{SU}}(8)$ representations. In this language there are two embedding tensor components corresponding to the two terms in the branching ${\bf{912}} = {\bf{36}} + {\bf{420}} + {\rm{c.c}}$
\begin{equation}
{\rm{\Theta}}_{\mathcal{M}}{}^{\alpha} \, \rightarrow \quad A^{IJ} \in {\bf{36}} \, , \quad A_{I}{}^{JKL} \in {\bf{420}} \quad \text{with} \; I,J,K,L \in {\bf{8}} \; \text{of} \; {\rm{SU}}(8) \, .
\end{equation}
The scalar potential (\ref{Scalar potential E7}) can be written in the elegant form
\begin{equation} \label{Scalar potential SU8}
V = - \dfrac{3}{8} \, A^{IJ} \, A_{IJ} + \dfrac{1}{24} \, A_{I}{}^{JKL} \, A^{I}{}_{JKL} \, .
\end{equation}
Notice that the nonlinear dependence on scalar fields has disappeared leaving us with a simple expression quadratic in the embedding tensor components. \\
Usually we are interested in finding critical points of (\ref{Scalar potential E7}) for a fixed gauging (thus for fixed embedding tensor components). Suppose for a given gauging ${\rm{\Theta}}_{0}$ we find a critical point $\vec{\phi} = \vec{\phi}_{0}$ for which $\partial_{\vec{\phi}} V(\vec{\phi}_{0}, {\rm{\Theta}}_{0}) = 0$. At the price of modifying the form of the embedding tensor we can bring the critical point to the origin of moduli space using a field independent  ${\rm{E}}_{7}$ transformation. This means that, leaving the embedding tensor free, every critical point could be brought to the origin. \\
This is exactly the philosophy of the GTTO approach. We can choose embedding tensor components which are free to take any possible configuration invariant under a particular subgroup ${\rm{G}}_{r} \subset {\rm{E}}_{7}$ and compatible with the LC and QC. At the origin of moduli space we compute the first derivative of the scalar potential. Being at the origin, the EOM are just a set of equations quadratic in the embedding tensor components. We solve these equations finding the set of structure constants which are compatible with having a critical point at the origin $\partial_{\vec{\phi}} V(0, {\rm{\Theta}}'_{0}) = 0$. \\
In this way we are sure to find all critical points of gauged maximal supergravity which have as residual gauge symmetry the group ${\rm{G}}_{r}$. Clearly the critical point we have found at the origin with the particular set of embedding tensor components ${\rm{\Theta}}'_{0}$ could be related to another critical point at the position $\vec{\phi} = \vec{\phi}_{0}$ but with a simpler set of structure constants ${\rm{\Theta}}_{0}$, but it is difficult to tackle this correspondence. In other words the drawback of this procedure is that, in principle, we have lost contact with the full gauge group ${\rm{G}}_{g}$.

\section{All ${\rm{G}}_2$ invariant critical points}
We are now ready to show how, using the GTTO approach, it is possible to classify all critical points preserving a certain amount of gauge symmetry. We fix the residual symmetry to be at least ${\rm{G}}_2$. This means the embedding tensor components need to be constructed using ${\rm{G}}_2$ singlets. \\
In order to parametrise ${\rm{G}}_2$ invariant tensors, we split indices according to
\begin{equation} \label{decomposition of the 8}
   I = \big( 1, m \big) \, , 
 \end{equation}
where $m,n,...$ is the fundamental of ${\rm{G}}_2$. The latter is also the fundamental of ${\rm{SO}}(7)$ when embedded in ${\rm{SO}}(8)$ in the standard way. ${\rm{G}}_2$ is defined as the subgroup of ${\rm{SO}}(7)$ that leaves invariant a particular 3-form $\varphi_{mnp}$ and its dual 4-form in seven dimensions. Decomposing the $\bf 36$ and $\bf 420$ of ${\rm{SU}}(8)$ into ${\rm{G}}_2$, one finds two and three singlets, respectively. We will parametrise these with the following Ansatz:
\begin{align}
A^{IJ} \, \rightarrow & \quad A^{11} = \alpha_1 \,, \quad A^{mn} = \alpha_2 \delta^{mn} \, , \\
A_{I}{}^{JKL} \, \rightarrow & \quad A_1{}^{mnp} = \beta_1 \varphi^{mnp} \,, \quad A_m{}^{1np} = \beta_2 \varphi_{m}{}^{np} \,, \quad A_{m}{}^{npq} = \beta_3 (* \varphi)_m{}^{npq} \, . \nonumber
\end{align}
Plugging the most general ansatz with five complex parameters into the QC and the EOM one gets a number of quadratic constraints on these parameters. As explained in more detail in \cite{Dibitetto:2011gm}, these are amenable to an exhaustive analysis by means of algebraic geometry techniques, in particular prime ideal decomposition, and the corresponding code \textsc{singular} \cite{DGPS}. In this way we find the four branches of solutions listed below, all corresponding to Anti-de Sitter space-times. In all cases we will omit an overall scaling of the solutions and use ${\rm{SU}}(8)$ to set the phases of $\alpha_{1}$ and $\alpha_{2}$ equal. They are either $\mathcal{N}= 0,1$ or $8$
\begin{itemize}
\item The first branch is $\mathcal{N} = 8$ and reads
\begin{equation}
\vec{\alpha} = \big(e^{i \, \theta}, e^{i \, \theta} \big) \, , \quad \vec{\beta} = \big(0, 0, 0 \big) \,.
\end{equation}
All solutions are ${\rm{SO}}(8)$ invariant and preserve $\mathcal{N} = 8$. They correspond to the origin of the standard ${\rm{SO}}(8)$ gauging and its one-parameter generalisation. As also noted in \cite{Dall'Agata:2012bb}, the mass spectrum is equal for the entire branch and given by
\begin{equation}
m^2 L^2 = - 2 \; (\times 70) \,,
\end{equation} 
\item The second branch is $\mathcal{N} = 1$ and is given by
\begin{equation}
\vec \alpha = \big(- 2 \, e^{- 5 i \, \theta}, \sqrt{6} \, e^{-5 i \, \theta} \big) \,, \quad \vec{\beta} = \big(0, \sqrt{2/3} \, e^{- i \, \theta}, e^{3 i \, \theta} \big) \, .
\end{equation}
For all values of these parameters, the invariance group is ${\rm{G}}_{r} = {\rm{G}}_2$ and the mass spectrum reads
\begin{equation}
m^2 L^2 =  (4 \pm \sqrt{6} ) \; (\times 1) , \quad 0 \; (\times 14) , \quad - \tfrac{1}{6} (11 \pm \sqrt{6}) \; (\times 27) \, .  
\end{equation}
This coincides with the ${\rm{G}}_2$ invariant mass spectra of the standard ${\rm{SO}}(8)$ theory. The latter corresponds to a particular value of $\theta$.  Other values include the one-parameter generalisation of \cite{Dall'Agata:2012bb} and possibly more. 
\item The third branch is $\mathcal{N} = 0$ and reads
\begin{equation}
\vec{\alpha} = \big(3 \, e^{-3 i \, \theta},  3 \, e^{-3 i \, \theta} \big) \,, \quad \vec{\beta} = \big(- e^{ i \, \theta}, e^{ i \, \theta}, \mp e^{i \, \theta} \big) \,.
 \end{equation}
The stability group in this case is ${\rm{G}}_{r} = {\rm{SO}}(7)_\pm$. The mass spectrum is independent of the parameter and reads
 \begin{equation}
 m^2 L^2 = 6 \; (\times 1) , \quad 0 \; (\times 7) , \quad - \tfrac{6}{5} \; (\times 35) , \quad - \tfrac{12}{5} \; (\times 27) \, .
 \end{equation}
The lowest of the eigenvalues violates the BF bound and hence this branch is pertrubatively unstable. The spectrum coincides with the ${\rm{SO}}(7)_\pm$ invariant mass spectra of the standard ${\rm{SO}}(8)$ theory, as was later explained group theoretically in \cite{Kodama:2012hu}. Again, the latter corresponds to a single point in a one-dimensional parameter space of non supersymmetric ${\rm{SO}}(7)_{\pm}$ invariant critical points. 
\item The last branch is $\mathcal{N}= 0 $ as well and is given by
\begin{equation}
\vec{\alpha} = \big( \sqrt{3} \, e^{- 3 i \, \theta}, - e^{- 3 i \, \theta} \big) \, , \quad \vec{\beta} = \big( e^{i \, \theta}, \tfrac13 \sqrt{3} \, e^{i \, \theta} \big) \,.
\end{equation}
The invariance group is ${\rm{G}}_{r} = {\rm{G}}_2$ and the mass spectrum reads
\begin{equation}
m^2 L^2 =  6 \; (\times 2) , \quad 0 \; (\times 14) , \quad - 1 \; (\times 54) \, .  
\end{equation} 
In this case all eigenvalues satisfy the BF bound, and hence this family of critical points is non-supersymmetric and nevertheless perturbatively stable. Previously known examples of stability without supersymmetry were isolated points with smaller symmetry groups \cite{Fischbacher:2010ec, Dibitetto:2012ia}. 
\end{itemize}
it is easy to compare the results in this section with the ones in the last section. We see that all critical points appearing in Fig.1 with the related symmetries and mass spectra fall in one of the listed branches. There is nontheless one fundamental difference. The $\omega$ parameter of section 3 describes theories having the same ${\rm{SO}}(8)$ gauge group. Within this one parameter family of theories we can follow the evolution of every critical point of the potential (\ref{potential G2}). In this section we are following a different evolution. For every branch the parameter $\theta$ describes the evolution of a critical point with fixed residual gauge symmetry and supersymmetry along the set of different theories which are compatible with it. \\
We will better clarify this point by analysing more specifically a particular branch, e.g. the ${\rm{SO}}(7)_{+}$ one. In order to show the physical changes when traversing the $\theta$ space we have calculated the eigenvalues of the Cartan-Killing metric, from which the full gauge group (and not only the invariance group of the critical point) can be derived. Again, this is outlined in \cite{Dibitetto:2011gm} and we employ the mapping given in \cite{Dibitetto:2012ia}. For the ${\rm{SO}}(7)_{+}$ critical points a set of 21 eigenvalues is always negative, corresponding to the preserved part of the gauge group. The remaining 7 are either all negative, zero or positive, as a function of $\theta$, leading to the following gauge groups:
\begin{align}
\theta & \in \big[ 0 , \arccos \sqrt{\tfrac{1}{6} ( 3 + \sqrt{5})} \, \big) \, : & {\rm{G}}_{g} & = {\rm{SO}}(7,1) , \nonumber \\
& = \arccos \sqrt{\tfrac{1}{6} ( 3 + \sqrt{5})} \, : & {\rm{G}}_{g} & = {\rm{ISO}}(7,1) , \nonumber \\
& \in \big( \arccos \sqrt{\tfrac{1}{6} ( 3 + \sqrt{5})} \, , \tfrac{\pi}{4} \big] \, : & {\rm{G}}_{g} & = {\rm{SO}}(8) . 
\end{align}
The gauge group therefore changes from compact to non-compact and viceversa, while passing trought an In\"on\"u-Wigner contracted point. \\
In this sense the GTTO approach contains the classical approach as the $\theta$ parameter describes more than just the ${\rm{SO}}(8)$ gauged theory critical points.

\section{Conclusion}
In this contribution we have briefly described some of the latest developments in the field of gauged maximal supergravity in four dimensions. After introducing the embedding tensor formulation of the theory, we have explained how the new one parameter family of ${\rm{SO}}(8)$ gauged supergravity has been discovered opening up a new landscape of theories. Finally we have shown how the GTTO approach could be used as a powerful tool for the study of vacuum structure. \\
Many questions are still waiting for an answer. First of all we need to understand whether this additional parameter is present for other gaugings, how big is the parameter space of new theories and what is the periodicity of this space. This would be of crucial importance providing new explicit examples to study with new interesting properties for many purposes. Related to this, it is still an open problem to identify the higher dimensional origin of the $\omega$ phase for the ${\rm{SO}}(8)$ gauged theory. Finally we need to understand how far we can go using the GTTO approach. In principle this approach allows to classify vacua with a generic residual symmetry. Unfortunately, the smaller the residual symmetry group the more expensive in terms of computational power is the task.

\section*{Acknowledgements}
AB would like to thank the organizers of the Barcelona Postgrad Encounters on Fundamental Physics for the beautiful conference and the warm hospitality. The research of AB and DR is supported by a VIDI grant from the Netherlands Organisation for Scientific Research (NWO). The work of AG is supported by the Swiss National Science Foundation.

\bibliography{Proceedings-arXiv}
\bibliographystyle{utphysmodb}

\end{document}